\pdfoutput=1

\documentclass[3p,twocolumn]{elsarticle}
     \makeatletter
\def\ps@pprintTitle{%
     \let\@oddhead\@empty
     \let\@evenhead\@empty
     \def\@oddfoot{\footnotesize\itshape
        \ifx\@journal\@empty 
       \else\@journal\fi\hfill 12 October 2017}%
     \let\@evenfoot\@oddfoot}
     \makeatother

\usepackage{amssymb,amsmath,amsthm, amsfonts}
\usepackage[font=small,labelfont=bf]{caption}
\usepackage{subcaption}
\usepackage{booktabs}
\usepackage{multirow}
\usepackage{enumitem}
\usepackage[T1]{fontenc}
\usepackage{mathtools}
\usepackage{hyperref}
\usepackage{graphicx}
\usepackage{rotating}
\usepackage{siunitx}
\setlist{nolistsep,leftmargin=*}

\begin{document}

\begin{frontmatter}



\title{A Survey on Behavioral Biometric Authentication on Smartphones}


\author[Minia]{Ahmed Mahfouz}
\ead{e.ahmedmahfouz@mu.edu.eg}
\author[Minia,CIC]{Tarek M. Mahmoud}
\ead{d.tarek@mu.edu.eg}
\author[Helwan,Sinia]{Ahmed Sharaf Eldin}
\ead{profase2000@yahoo.com}

\address[Minia]{Computer Science Department, Minia University, El-Minia, Egypt}
\address[CIC]{Canadian International College (CIC), Cairo, Egypt}
\address[Helwan]{Information Systems Department, Helwan University, Egypt}
\address[Sinia]{Faculty of Information Technology and Computer Science, Sinai University, Egypt}

\cortext[cor]{Corresponding author.}

\begin{abstract}
Recent research has shown the possibility of using smartphones' sensors and accessories to extract some behavioral attributes such as touch dynamics, keystroke dynamics and gait recognition. These attributes are known as behavioral biometrics and could be used to verify or identify users implicitly and continuously on smartphones. The authentication systems that have been built based on these behavioral biometric traits are known as active or continuous authentication systems.

This paper provides a review of the active authentication systems. We present the components and the operating process of the active authentication systems in general, followed by an overview of the state-of-the-art behavioral biometric traits that used to develop an active authentication systems and their evaluation on smartphones. We discuss the issues, strengths and limitations that associated with each behavioral biometric trait. Also, we introduce a comparative summary between them. Finally, challenges and open research problems are presented in this research field.
\end{abstract}


\end{frontmatter}

\begin{figure*}
    \centering
    \includegraphics[scale=0.25]{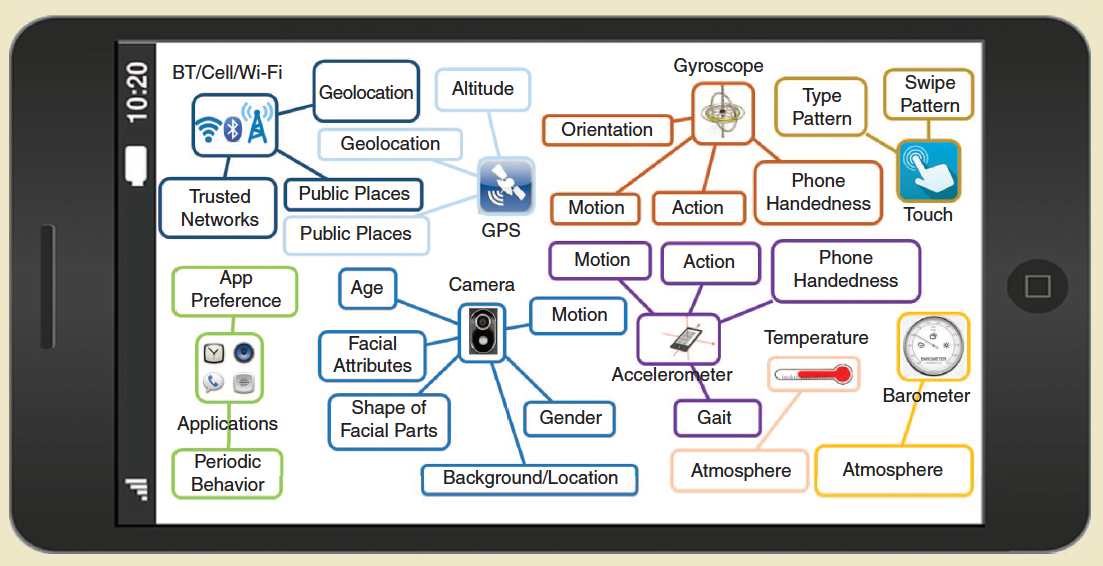}
    \caption{Sensors, services and devices in smartphones that could be used to make a person recognition based on physiological and behavioral traits~\cite{Patel:2016}.}
    \label{fig:biometric_sensors}
\end{figure*}
\section{Introduction}\label{introduction}

With the diversity of sensors and services on smartphone as shown in Figure~\ref{fig:biometric_sensors}, the smartphone became more smarter and attracts both (1) users who enjoy using it to facilitate their daily life more than ever before, consequently they store more sensitive and private information on it, and (2) attackers who pay more attention to access or steal these sensitive data. These attacks could be done by either \textbf{insider attacker}, someone who know the user such as friend or family member or \textbf{stranger attacker}, someone who does not know the user~\cite{Muslukhov:2012}.

Due to the weaknesses of the traditional authentication mechanisms such as PIN, Pattern and Password, and the biometric based mechanisms such as fingerprint, face and voice recognition on smartphones, the research community have developed authentication mechanisms based on behavioral biometric traits such as gesture, keystroke and gait. These mechanisms are known as active or continuous authentication mechanisms.

In this paper we present the components and the operating process of the active authentication mechanisms in general, followed by some different metrics that used to evaluate the performance of an active authentication mechanisms. We also conducted an extensive survey of the state-of-the-art active authentication systems and their evaluation on smartphones. We discuss the issues, strengths and limitations that associated with each behavioral biometric trait, and introduce a comparative analysis between them. Finally, we identify challenges, open research problems and provide a set of recommendations in this research field.

The rest of the paper is organized as follows: Section~\ref{active_authentication_components} provides an overview of active authentication systems in general. Section~\ref{biometric_trait_selection} presents a set of factors that facilitate the selection of a behavioral biometric trait. Section~\ref{design_factor} presents another set of factors that help in the designing process of the biometric authentication system. Section~\ref{common_traits} surveys the common behavioral biometric traits. Section~\ref{limitations} presents some limitations and followed by set of challenges and future trends.
 
 \begin{figure*}
    \centering
    \includegraphics[scale=0.45]{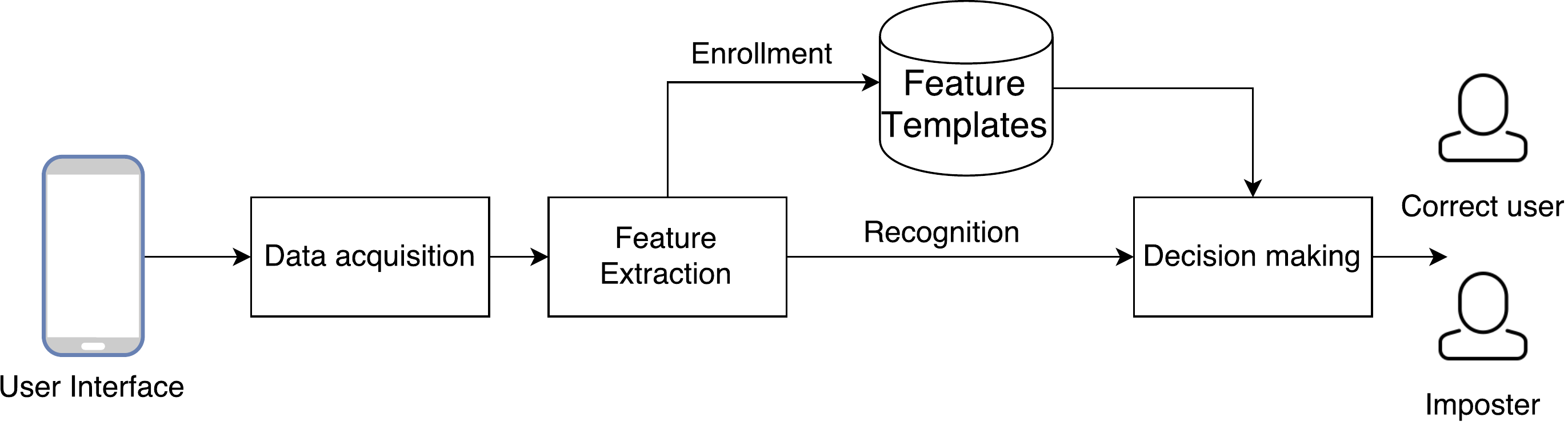}
    \caption{The operation of a biometric recognition system.}
    \label{fig:biometric_operation}
\end{figure*}

\section{Adversary Attacks}
The main goal of attackers is to gain physical access to the victim's device for snooping or data destruction. These attacks could be done by either \textbf{insider attacker} or \textbf{stranger attacker}~\cite{Muslukhov:2012}.\\\\
\textbf{Insider Attacker}, someone who know the user such as friend or family member. The insider attacker has opportunity to have unauthorized access to the victim's device due to the proximity between them. Based on a previous research done by Usmani et al.~\cite{Usmani:2017} where they characterized the social insider attacks and found that the existing devices ( i.e., which use the traditional authentication methods such as Pattern or Password) and the Facebook account security measures are ineffective to resist social insider attacks.\\\\
\textbf{Stranger Attacker}, someone who does not know the user. The stranger attacker has no prior knowledge about the victim, who may steal the legitimate user's device or found a lost device.

\section{The Active Authentication}\label{active_authentication_components}
In this section we define what is an active authentication and show an overview on how does the active authentication system work, followed by its modes of operation. Finally, we present different metrics that have been used to evaluate the performance of active authentication systems. 
\subsection{What is an active authentication system?}
Active authentication system is an automated recognition process that verifies or identify individuals based on detailed information about their body such as face or their behaviors such as how they type or interact with some sensors on smartphone. Figure~\ref{fig:biometric_sensors} shows some sensors and services that can be used to acquire behavioral biometric data. The main goal of the recognition process is to prevent the unauthorized access from imposters and grant access only for legitimate user. The idea behind the recognition process in active authentication system is to establish an identity based on \textbf{who you are?} concept. The details of how recognition process work based on a specific biometric trait will be described in the next section.

There are two important characteristics that should be achieved by any active authentication system which are as follows:
\begin{itemize}
\item \textbf{Continuity}: A smartphone verifying user in a continuous manner, where the authentication system keep authenticating users as long as the user uses the smartphone. In other words, it is a re-authentication process that conducts periodically.
\item \textbf{Transparency}: All authentication processes should be carried out in the background without interrupting the user (i.e., user will be implicitly authenticated without any intervention).
\end{itemize}

The two aforementioned characteristics are representing the cornerstone of any active authentication system, which make it different than the traditional authentication system. There are different biometric techniques could be used to achieve these characteristics. These techniques are categorized into two groups, physiological biometric mechanisms such as face and voice recognition, and behavioral biometric mechanisms such as touch and keystroke dynamics. In this paper we concentrate on surveying the behavioral biometric ones.

\subsection{How does the active authentication system work?}\label{operation}
The active authentication system works similarly like the biometric recognition system, which contains two main phases, enrolment phase and recognition phase as shown in Figure~\ref{fig:biometric_operation}. In the enrolment phase, the system acquires the biometric data, analyzes this data and extracts a distinctive features set, then it builds the feature templates (e.g., like the training process for a classifier). In the recognition phase, the system, similarly, acquires biometric data and extracts features, but instead of storing these features in the feature templates, it compares it with the stored one to verify the user identity.

There is a set of basic modules should be included in any active authentication system in general which are as follows:
\begin{enumerate}
\item \textbf{Data acquisition module}: it is the first step in the system where the raw biometric data is collected by one of the sensors in the smartphone such as camera or touchscreen sensor (see Figure~\ref{fig:biometric_sensors}). The quality of the collected data is very important because it will affect on the successor modules of the recognition process. The quality of data is impacted by the used sensors and the environment in which the data was collected~\cite{Jain:2011:IB:2161587}. 
\item \textbf{Feature extraction module}: before extracting the distinctive features, the raw data has to be preprocessed, detect and remove outliers, improve the data quality, especially if the data collected in an uncontrolled environment with uncooperative users. Then, once the data is cleaned and processed, set of discriminative features are extracted. The extracted features depend on the type of raw data, for example if the collected data contains timestamps, temporal feature could be extracted.
\item \textbf{Feature templates}: it is a repository database that contains a concatenation of the extracted feature vectors for a specific user (i.e., device owner). It is built during the enrollment phase and used during the recognition phase to be compared with the captured feature sample to verify the claimed identity.
\item \textbf{Matching and decision-making module}: it used only during the recognition process, where it compares the extracted features against the stored feature templates to generate a matching score to make a decision. The decision validates the claimed identity to see it is done by legitimate user or imposter.
\end{enumerate}

\subsection{Modes of operation}
There are two different modes that the biometric system could operate, which depends on the recognition context mode, verification or identification. 
\subsubsection{Verification mode}
In verification mode, which is one-to-one matching process, the system verifies the claimed identity by comparing it with the stored one. If the matching score of the claimed identity greater than a predefined threshold $\alpha \in (0,1)$, then the claimed identity is accepted as legitimate, otherwise, the claimed identity is rejected as imposter. So, authentication process could be operate based on verification mode and implemented as a binary classification problem. The decision rule is calculated based on the following formula:
\begin{equation}
p(u_i) =
  \begin{cases}
    legitimate       & \quad \text{if } p(u_i) > \alpha\\
    imposter  & \quad \text{if } p(u_i) < \alpha\\
  \end{cases}
\end{equation}
where $p(u_i)$ represents the authentication score for a user $u_i$ and is calculated by the classifier, and $\alpha$ represents a predefined threshold $\in(0, 1)$.
\subsubsection{Identification mode}
In identification mode, which is one-to-many matching process, the system recognizes the presented biometric sample by comparing it with all stored templates (i.e., a template for each user), where the matching algorithm estimates the identity of the sample based on the highest matching score and a designated threshold (i.e., there is multiple matching scores will be generated, one for each user, in which the highest score will be selected).

\subsection{Performance metrics}\label{performance_metrics}
There are different metrics could be used to evaluate the performance of an active authentication system. Selecting metrics depends on the type of evaluation, and there are three types of evaluation could be performed:
\begin{enumerate}
\item \textbf{Technology evaluation}: is the dominant general type of evaluation testing. It is used to evaluate the same biometric modality in offline mode and compares different algorithms within a single modality on a fixed dataset. 
\item \textbf{Scenario evaluation}: the main objective of this evaluation type is to test the whole biometric system for a class of applications in a real world manner where the dataset collected from real subjects.
\item \textbf{Operational evaluation}: it is similar to scenario evaluation but it measures a comprehensive biometric system in specific application environment in a real-time manner.
\end{enumerate}

Because our application context here is authentication, we describe the metrics that could be used in verification mode rather than identification mode. Our assumption is that each mobile device is used only by one user (i.e., single user context) and our goal is to prevent the unauthorized access by differentiating between the legitimate user and imposter (i.e., binary-class classification problem).

The basic metrics that used to evaluate the active authentication system are depending on the error rates. Before describing the verification system error rates, we are going to mention some basic metrics that will be used in verification error calculation.

\begin{figure}
    \centering
    \includegraphics[scale=.4]{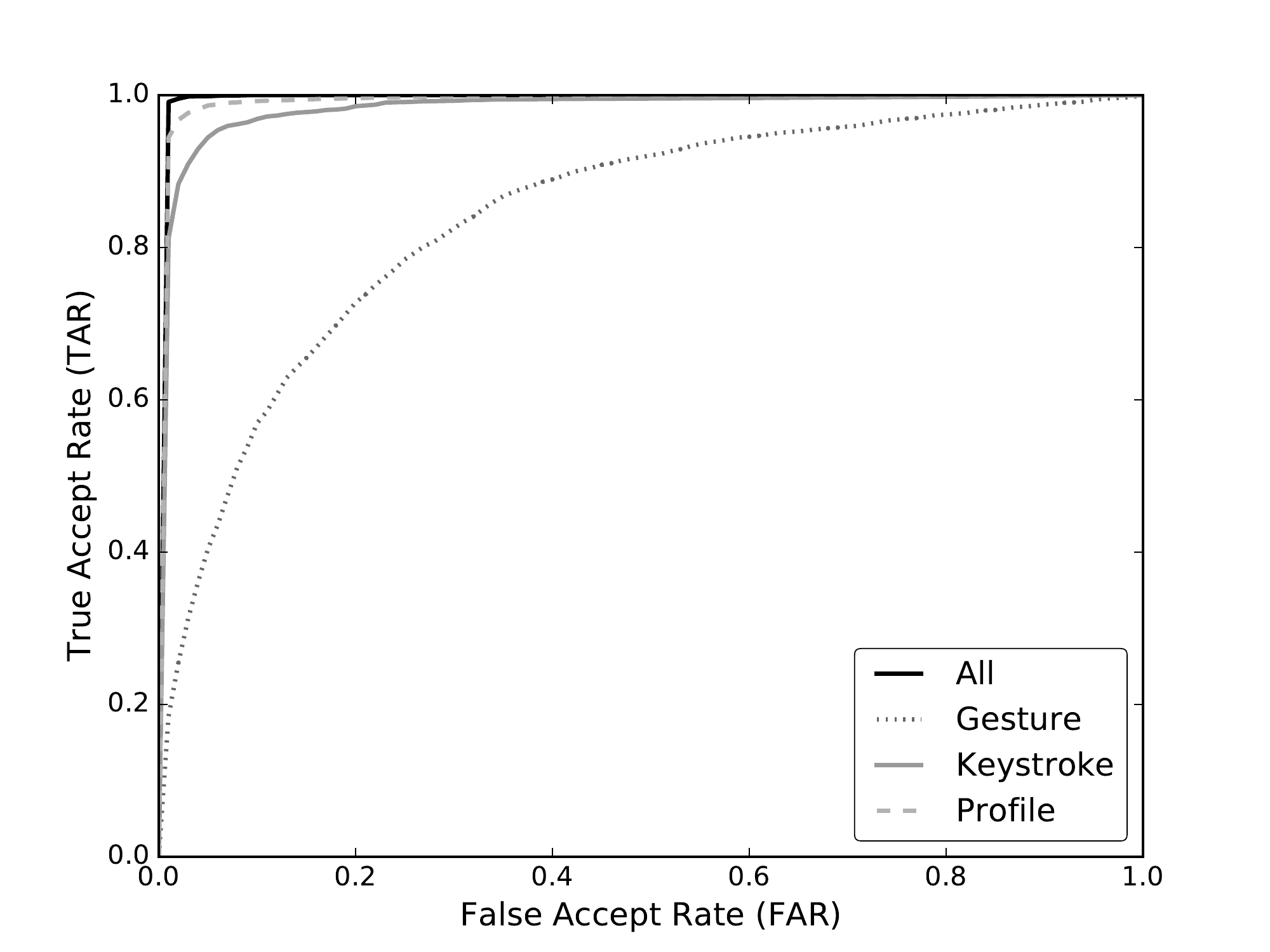}
    \caption{The performance of a biometric system can be summarized using Receiver Operating Characteristic (ROC). In this figure, the performance curves are computed using different biometric traits (Gesture, Keystroke, Profile sensors). ROC curves plots TAR along the y-axis and FAR along the x-axis. The area under the curve (AUC) is used to quantify the quality of the authentication model as an alternative to the accuracy.}
    \label{fig:sample_ROC}
\end{figure}

The raw basic metrics and their description in our problem domain are as follows:
\begin{itemize}
\item \textbf{True Accept (TA)}: The system correctly matches a genuine user to the corresponding template stored within the system.
\item \textbf{True Reject (TR)}: The system correctly denies an imposter, where its data that not matching to any template within the system.
\item \textbf{False Accept (FA)}: The impostor was incorrectly matched to a genuine user template stored within a biometric system.
\item \textbf{False Reject (FR)}: The genuine user is incorrectly rejected from the system.
\end{itemize}

The common metrics that have been used in the literature to evaluate the performance of the active authentication system are as follows: 
\begin{itemize}
\item \textbf{True Accept Rate (TAR)} describes the probability that the system correctly matches a genuine user to the corresponding stored template within the system, and is calculated based on the following formula:
\begin{equation}
TAR = \frac{TA}{TA + FR}
\end{equation}
\item \textbf{False Accept Rate (FAR)} describes the proportion of impostors that were incorrectly matched to a genuine users templates stored in a biometric system, and is calculated based on the following formula:
\begin{equation}
FAR = \frac{FA}{FA + TR}
\end{equation}
FAR reflects the ability of a non-authorized user to access the system, whether via zero-effort access attempts or deliberate spoofing or any other method of circumvention.
\item \textbf{False Reject Rate (FRR)} describes the proportion of genuine users that were incorrectly rejected from a biometric system, and is calculated based on the following formula:
\begin{equation}
FRR = \frac{FR}{TA + FR}
\end{equation}
\item \textbf{Equal Error Rate (EER)}: describes the point at which genuine and imposter error rates are equal, where the lower EER indicates better performance. It could be used to summarize the performance of the authentication system in a single value result. Previous research has been conducted to calculate it with respect to energy consumption. For instance, Sitova et al.,~\cite{Sitova:2013} proposed an evaluation for the active behavioral biometric authentication system based on EERs with respect to various levels of energy consumption. 
\end{itemize}

Also, there are two important metrics that could be used to describe the authentication performance of the system in a presented curve:
\begin{itemize}
\item \textbf{Receiver Operating Characteristic (ROC) Curve}: depicts the trade-off between TAR~along the y-axis and FAR~along the x-axis in a single curve at various threshold values, where points are plotted parametrically as a function of the decision threshold~(see Figure~\ref{fig:sample_ROC}). The top left corner of the plot represents the ideal point, where TPR equal one and FPR equal zero.
\item \textbf{Area Under Curve (AUC)}: it is used to quantify the quality of the authentication model as an alternative to the accuracy (see Figure ~\ref{fig:sample_ROC}). Also it is useful even when there is a high class imbalance (i.e. one of the classes dominates). The value of AUC ranges from 0.5 to 1, where 0.5 represents the random guessing and 1 represents the ideal results (i.e., no errors in the system).
\end{itemize}

\section{The Selection of The Biometric Trait}\label{biometric_trait_selection}
There are different biometric traits could be used in active authentication system which relies on the application context. Each trait could be used in certain context but others not. There are some factors could be used to evaluate the suitability of the biometric trait which are as follows~\cite{Jain:1998}:
\begin{itemize}
\item	\textbf{Universality}: each user should have the biometric trait.
\item	\textbf{Uniqueness}: the biometric trait should be adequately differentiates between any two users. This will help to generate a discriminative features that could be used to differentiate between legitimate user and imposter with high accuracy.
\item	\textbf{Permanence}: the biometric trait should be durable (i.e., not vary over time). 
\item	\textbf{Collectability}: the biometric trait should be easy to collect and measure.
\item	\textbf{Performance}: the accuracy of the biometric trait should be robust and functional for the given environment.
\item	\textbf{Acceptability}: the users should accept and willing to present her biometric trait.
\item	\textbf{Circumvention}: the biometric trait should not be susceptible to spoofing or any other attacks.
\end{itemize}

\section{Design Factors}\label{design_factor}
The design of the biometric authentication system is influenced by some factors~\cite{Jain:2011:IB:2161587}, we describe them in regard to mobile application environment which are as follows:
\begin{itemize}
\item	\textbf{User cooperation}: Cooperation refers to the behavior of the user when interacting with the authentication system, where the biometric trait is collected either from cooperative user, where the user interacts with the system in concerted manner, or uncooperative user, where the user does not perform the trait as it should be.
\item	\textbf{The degree of control}: the degree to which a deployment environment is controlled or uncontrolled; whether the deployment environment is outdoors, indoors, or mixed.
\item	\textbf{User awareness}: explores if user is aware that he is being subjected to biometric recognition system or not.
\item	\textbf{The habituation}: explores if the user has experience to interact with the biometric system before or not.
\item	\textbf{Open versus closed system}: explores If a person's biometric template can be used across multiple applications, in this case the system is open, otherwise, the system is closed.
\end{itemize}

\section{Common Behavioral Biometric Traits}\label{common_traits}
\begin{figure*}[!t]
    \centering
    \includegraphics[scale=0.35]{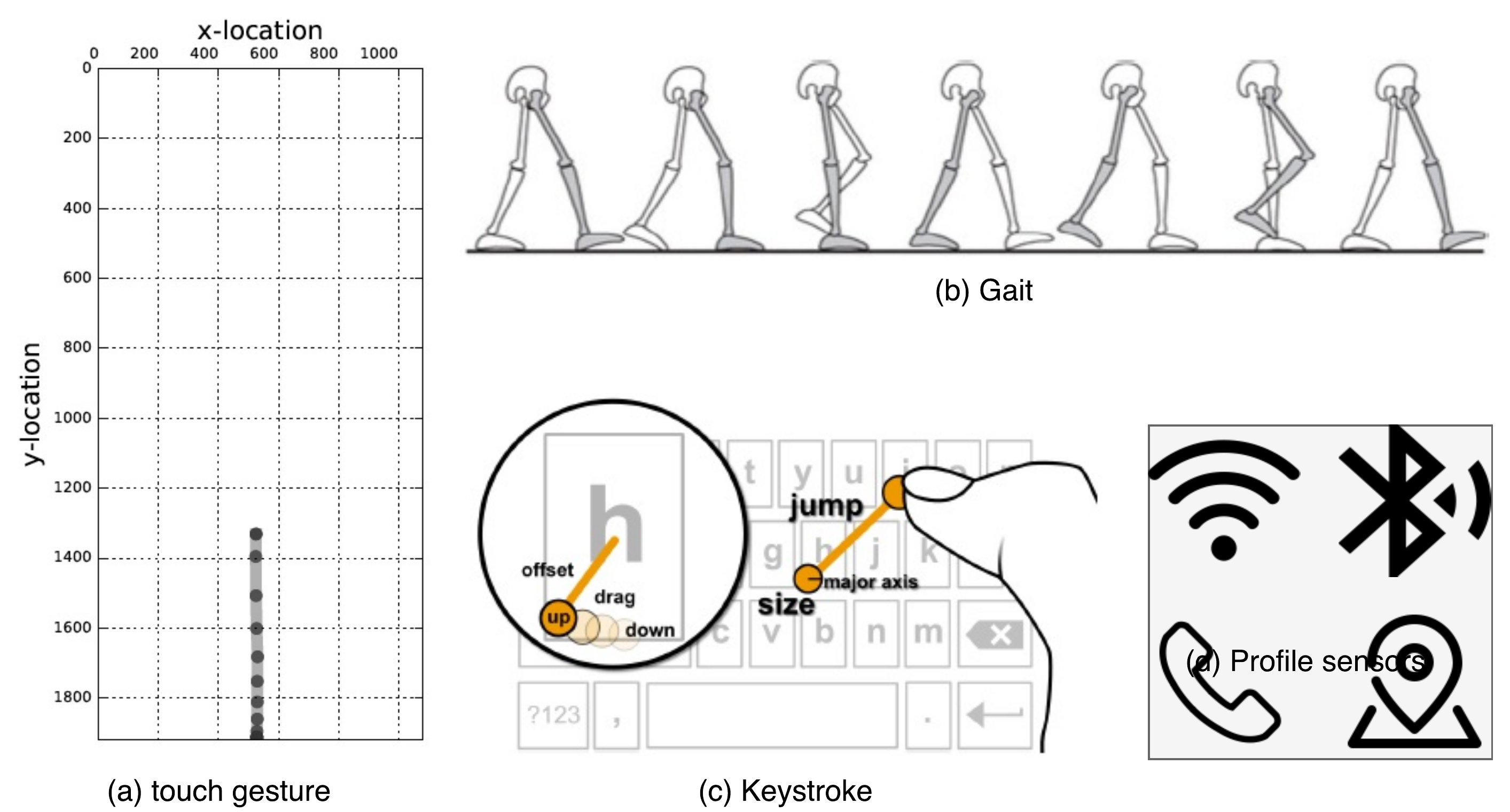}
    \caption{Examples of behavioral biometric traits that have been used for active authentication. It includes (a) touch gesture, (b) gait~\cite{hoang:2015}, (c) keystroke dynamics~\cite{buschek:2015} and (d) behavioral profiling.}
    \label{fig:behavioral_traits}
\end{figure*}

Behavioral biometric trait is a particular characteristics that can be acquired from user actions such as touch gesture, keystroke dynamics or behavioral profiling~(See Figure~\ref{fig:behavioral_traits}). In this section we present a review about some of the commonly used behavioral biometric traits that have been proposed in the literature to design an active authentication systems on smartphones. 

\subsection{Gesture based authentication}
All gesture based authentication methods are built based on the analysis and measuring of touch gestures on smartphones~\cite{gesture:2016}. The touch gesture biometric trait is a hand drawn shape on the smartphone touch screen that contains one or more strokes. The stroke is a sequence of consecutive timed points. Each point represented by an ordered pair of numerical coordinates~as shown in Figure~\ref{fig:behavioral_traits}a. Smartphone touch screen represents the input data source for gesture based authentication, which is the main input method used in the smartphones. Touch gestures could be acquired from application level~\cite{motionevent:2017} or operating system level~\cite{motioneventOS:2017}. Every touch gesture includes touch mechanics which depicts what your fingers do on the screen?, and touch activities which depicts the results of specific gestures~\cite{gesture:2016}.

A useful discriminative set of features could be extracted from the touch gesture biometric trait, but at the same time, it also introduces some challenges. We are going to discuss how touch gesture biometric trait has been used to develop an active authentication system.

\begin{table*}
\centering
\scalebox{0.75}{
\begin{tabular} {l|p{1.65cm}| l | l | l | l | l} 
\toprule 
\textbf{Study} & \textbf{\# of Participants}  & \textbf{Dataset (\# of samples)} & \textbf{Platform} & \textbf{\# of features} & \textbf{Classification} & \textbf{Performance(\%)} \\ \hline
Shahzad et al.~\cite{Shahzad:2013}&50&Private (15009 overall)&Windows&7&SVDE&EER:0.5\\
Zhao et al.,~\cite{Zhao:2013}&30&Private(120/user)&Android&Image (100x150)&NC,L1,L2 distance&EER:2.62\\
Serwadda et al.~\cite{serwadda:2013}&190&Private(50/touch type)&Android&28&10 Classifiers&EER:10.5-42.0\\
Xu et al.~\cite{Xu:2014}&42&Private(200/user)&Android&(4,37,42,49)&SVM&EER:$<$10.0\\
Feng et al.~\cite{Feng:2012}&40&Private(-)&Android&53&DT, RF, BN&FAR:4.66, FRR:0.13\\
Frank et al.~\cite{Frank:2013}&41&Private(-)&Android&27&KNN, SVM&EER:$<$4.0\\
Li et al.~\cite{li:2013}&75&Private(400/user)&Android&10&SVM&EER~3.0\\
\bottomrule
\end{tabular}}
\caption{Gesture based authentication methods. It contains the-state-of-the-art studies that conducted to develop touch-based active authentication systems. Number of Participants depicts how many participants contributed to the study. Dataset contains status of the collected data, private or public. The platform OS in where the study conducted. Number of features that extracted from the collected data. Classifiers that used in the classification process. Finally, the best results that achieved by the developed method, where some studies evaluated based on EER and others based on FAR and FRR.}
\label{table:gesture_methods}
\end{table*}

\subsubsection{Data collection}
Data collection is the first step in the active authentication system, and the raw touch data is acquired from the touchscreen sensor. The behavior of touch gesture techniques is determined by a transfer function~\cite{Quinn:2012} that converts human input actions attached with some parameters such as (\textit{size, length, speed, velocity, pressure, or direction}) into a  gesture output effect. Most of proposed gesture authentication system in the literature are based on the assumption that the users are going to perform the touch gestures in away that reflects their behaviors. Hence, the parameters that attached with the touch input vary from one user to another. 

Touch gesture biometric trait has been used in active authentication systems because it implies two important characteristics:
\begin{itemize}
\item \textbf{Continuity}: Touch gesture biometric trait can be used to continuously authenticate users by monitoring their touch dynamics patterns. Users can be re-authenticated as long as they are using the smartphone. This is one of the most important advantages that touch gesture biometric trait has. Specially when compare it with the traditional authentication methods and physiological biometrics.
\item \textbf{Transparency}: Touch gesture biometric trait could be acquired without any interruption to the user. This is because the acquiring and processing of touch gestures can be carried out in the background while the smartphone being used by the user.
\end{itemize}

As illustrated in Table~\ref{table:gesture_methods}, the majority of conducted studies have collected data from less than or equal 50 participants~\cite{Shahzad:2013, Zhao:2013, Xu:2014, Feng:2012, Frank:2013}. On the other hand, few studies have conducted with a larger number of participants as in~\cite{serwadda:2013, li:2013}. The majority of formulated datasets contain hundreds of samples per user and were collected during a lab study from cooperative users in controlled environment~\cite{Teh:2016}. 

Regarding to the platform that used during data collection process, Android was the most common platform used for touch gesture biometric acquisition, due to its popularity in the market shares and it is easier to customize than iOS or Windows.

\subsubsection{Feature extraction}\label{gesture_feature}
Feature extraction is one of the main modules used in active authentication system, where the classifier classifies the users based on features set. The main goal of feature extraction is to identify and extract discriminative set of features by analyzing the raw touch gesture data from the users. The common extracted features were belonged to the following categories:
\begin{itemize}
\item \textbf{Temporal features}: one of the common used features set in touch gesture authentication system. Extracting this features set relies on the time analysis of user touches, where every touch gesture event is attached with timestamp. For example, the total time taken to perform a touch gesture could be calculated based on the difference between touch down and touch up timestamps, and we can use that calculated duration as a temporal feature.
\item \textbf{Spatial features}: extracted by doing analysis relating to the position, area, and size of the touch gesture, where every touch gesture is performed in specific position on the touch screen and represented by $x, y$ coordinates. Also touch size can be used which represents the approximation of the screen area that is being touched during a touch event. Another spatial feature which is touch pressure, a value measures the approximated force asserted on the screen for each touch event.
\item \textbf{Dynamic features}: extracted from the dynamic analysis of the touch event. For example the touch gesture is detected by the motion of object (i.e., finger) on the touch screen. By analyzing this motion helps to generate a useful set of features that could be used to differentiate between users. 
\item \textbf{Geometric features}: extracted by conducting a geometric analysis on the touch gesture. As the touch gesture contains one or more strokes and the stroke is a sequence of consecutive timed points. Analyzing the relationships between points, lines, curves that generated by conducting touch gestures give a useful discriminative features.
\end{itemize}

\subsubsection{Evaluation}
The evaluation of touch gesture biometric trait in the literature is based on verification mode (i.e., one-to-one matching)~\cite{Phillips:2000}, where the user claim the identity and the system validates the claimed presented identity. The following are set of components that affect on the evaluation process:
\begin{itemize}
\item \textbf{Dataset}: The dataset sample size has a huge impact on the accuracy of any proposed authentication method. Based on the literature, the majority of conducted studies collected small  samples (i.e., hundreds per user as shown in Table~\ref{table:gesture_methods}) such as in~\cite{Zhao:2013, serwadda:2013, Xu:2014, li:2013}. On the other hand, there are few studies that collected thousands of samples such as in~\cite{Shahzad:2013}. Evaluating the performance of a proposed active authentication system on large-scale dataset that contains millions of samples is going to be more realistic than using small-size datasets.
\item \textbf{Classification model}: different classification methods have been used in the touch gesture authentication. Some of them were based on probabilistic modelling such as Bayesian Network~\cite{Feng:2012}, and others used Support Vector Machine (SVM), which is another classification technique that separates the feature space by a hyperplane such that the margin between the two classes is maximized~\cite{li:2013, Frank:2013}. One of the most common used classification technique in touch gesture authentication systems is K-nearest-neighbors (KNN) which is robust and fast classification method that takes every new observation and locates it in feature space with respect to all training observations. Also, Decision Tree has been used to classify data based on the learned touch patterns as in~\cite{Feng:2012}. 
\item \textbf{Static and dynamic modes}: In the static mode, the identity of a subject is verified based on the input provided by the subject on the first instance of accessing a system. In the dynamic mode, a subject's identity is continuously verified throughout the active session of a mobile device.
\item \textbf{Metrics}: Equal Error Rate (EER) is the most common used metric in the literature to evaluate the performance of the touch gesture authentication. Table~\ref{table:gesture_methods} shows a comparison between the performance of different proposed active authentication systems. As we can see the most proposed systems achieve low EER values based on their collected datasets which are less than 10\%. Some proposed systems use other metrics such as Area Under The CurveAUC~\cite{Shahzad:2013}, False Rejection Rate (FRR) and False Acceptance Rate (FAR)~\cite{Feng:2012}, for more declaration about the most common used metrics see section~\ref{performance_metrics}.
\end{itemize}

\begin{figure}[t]
    \centering
    \includegraphics[width=\linewidth]{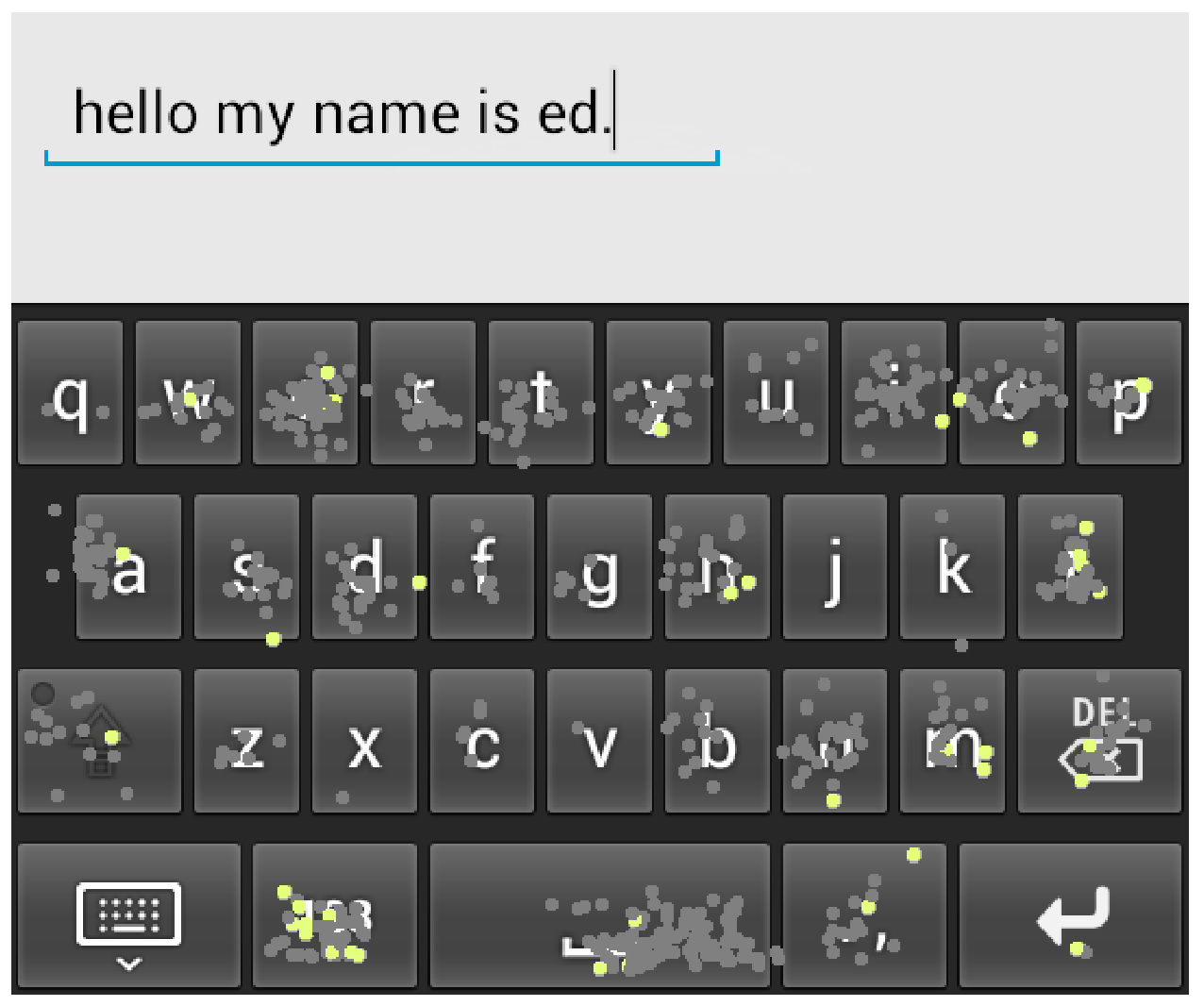}
    \caption{Screenshot from the virtual soft keyboard comparing user typing patterns. It shows the presses different locations done by two users, the normal user (grey dots) and other user (yellow dots). Image obtained from KeySens~\cite{Draffin:2014}.}
    \label{fig:keystrokes_keyboard}
\end{figure}

\begin{table*}[!h]
\centering
\scalebox{0.8}{
\begin{tabular} {l|p{1.8cm}| p{2.5cm} | l | p{1.8cm} | l | l} 
\toprule 
\textbf{Study} & \textbf{\# of Subjects}  & \textbf{Dataset (\# of sample)} & \textbf{Platform} & \textbf{\# of features} & \textbf{Classification} & \textbf{Performance(\%)} \\ \hline
Buschek et al.~\cite{buschek:2015}&28&Private(20160 overall)&Android&24&Knn,SVM,NB,LSAD&EER:26.4-36.8\\
Draffin et al.~\cite{Draffin:2014}&13&Private(430000)&Android&6&Neural Networks&FAR:14, FRR:2.2\\
Trojahn et al.~\cite{trojahn:2013}&18&Private(1980)&Android&32&J48 Decision Tree&FAR:2.03, FRR:2.67\\
Feng et al.~\cite{Feng:2013}&40&Private(-)&Android&122&J48,RF,BN&FAR:1.0,FRR1.0\\
Clarke et al.~\cite{Clarke:2006}&30&Private(30)&-&-&GRNN, RBF, FF MLP&EER:12.8\\
Gunetti et al.~\cite{gunetti:2005}&205&Private(-)&HTML form&-&Mathematical Model&FAR:5.0,FRR:0.5\\
\bottomrule
\end{tabular}}
\caption{Keystroke based authentication methods. It contains the-state-of-the-art studies that conducted to develop Keystroke-based active authentication systems. Number of subjects depicts how many participants contributed to the study. Dataset contains status of the collected data, private or public. The platform OS in where the study conducted. Number of features that extracted from the collected data. Classifiers that used in the classification process. Finally, the best results that achieved by the developed method. Abbreviation used: Least Squares Anomaly Detection (LSAD), Naive Bayes (NB), Support Vector Machine (SVM), J48 Decision Tree (J48), Random Forest (RF), Bayes Net (BN), k-Nearest Neighbors (k-NN), Equal Error Rate (EER), False Rejection Rate (FRR) and False Acceptance Rate (FAR), Generalized regression neural networks (GRNNs), Radial basis function (RBF), Feed forward multi-layered perceptrons (FF MLPs)}
\label{table:keystrokes}
\end{table*}
\subsection{Keystroke dynamics}
Keystroke dynamics is one of the old behavioral biometric trait that have been proposed for a long time to authenticate user continuously on computers~\cite{Monrose:2000, Bergadano:2002, gunetti:2005}. For smartphones, keystroke dynamics is the process of analyzing the way a user types on smartphone virtual keyboard by monitoring the keyboard inputs as shown in Figure~\ref{fig:keystrokes_keyboard}, and attempts to identify them based on habitual rhythm patterns in the way they type. With the diversity of touchscreen smartphones, the way a user types on smartphone has changed to be easier and more friendly. At the same time the raw data that associated with keystroke analysis became boarder and opened opportunity to collect more data and extract set of discriminative features that can be used to authenticate smartphone user based on the way they type~(keystroke dynamics). Based on the literature, we present how the data that relates to keystroke dynamics collected, followed by a description on extracted features set. Then, we discuss the evaluation techniques that have been used in the literature to evaluate the performance of the active authentication methods that built based on keystroke dynamics.

\subsubsection{Data collection}
Most research in the field of keystroke analysis collect data from structured and predefined text. For instance, Clarke et al.~\cite{Clarke:2006} asked participants to enter 30 text messages over three sessions. The messages contained mixture of quotes, lines from movies and typical text messages. Length of messages varied with average 14 words per message. Similarly, Feng et al.~\cite{Feng:2013} developed an android application to collect keystrokes over login session, where user asked to enter passwords (i.e., length is 4, 20 different passwords were used), and post-login session, where the user asked to enter a predefined sentences (i.e., the length of sentences varies from 14 words to 53 words, and on average 23 words). Trojahn et al.~\cite{trojahn:2013} also asked users to enter a specified sentence with 11 characters ten times (two words with one space in between). Buschek et al.~\cite{buschek:2015} invited participants to spend two sessions, with a gap of at least one week. Each session comprised three main tasks where the participants typed 6 different passwords in random order, 20 times each. As we can see, all aforementioned techniques relies on a specific context with a predefined text.
 
On the other hand, few studies have been conducted to collect data over all usage context i.e., not restricted with some predefined sentences or password. Draffin et al.~\cite{Draffin:2014} conducted a real-world field study to collect keystrokes from 13 users over three weeks period. They collected 86000 keypresses overall context without any intervention, not just from passwords or controlled phrases.

Similar to touch gesture, the keystroke dynamics can be used in active authentication systems because it implies two important characteristics:
\begin{itemize}
\item \textbf{Continuity}: keystroke dynamics can be used to continuously authenticate smartphone users by monitoring their way of typing, where the users can be re-authenticated as long as they type on the virtual keyboard. This is one of the most important advantages that keystroke dynamics has, specially when compare it with the physiological biometric traits.
\item \textbf{Transparency}: keystroke dynamics could be acquired implicitly while the users type without any interruption to them. This is because the acquiring and processing of touch gestures can be conducted in the background while the smartphone being used by the user.
\end{itemize}

\subsubsection{Extracted features}
Different features can be extracted from keystroke analysis. The most common used features in the literature of keystroke analysis are the duration and latency of the keypresses. The duration represents the amount of time between press and release of a key, and the latency is calculated based on the elapsed time difference between the release of the previous key and the press of the current key. Some other features are relevant to touch gesture as described in section~\ref{gesture_feature} such as spatial features (i.e., features relate to the position, area, pressure and size of the keystroke presses).

\subsubsection{Evaluation}
Most of the evaluation methods in the literature are based on verification mode such as in~\cite{maiorana:2011, buschek:2015, Feng:2013}, some other little evaluation methods have been used based on identification mode such as in~\cite{Nauman:2013}. 

Table~\ref{table:keystrokes} compares different proposed active authentication methods based on keystroke dynamics. Buschek et al.~\cite{buschek:2015} evaluated their proposed authentication mechanism on a private dataset with 20160 samples. They used different models to make verification. First they used classification models such as KNN, NB and SVM. Second, they used anomaly detection model such as LSAD~\cite{quinn:2014} and they found that the classification models performed better than anomaly detection model. Overall the system they reduced EER by 26.4 - 36.8\%. Trojahn et al.~\cite{trojahn:2013} formulated a dataset of 1980 samples collected from 18 users. They used different classification algorithm such as J48 decision tree, MLP, BayesNet and Naive Bayes. The best result for FAR and FRR error rates achieved by the J48 classifiers which were 2.03\% and 2.67\% respectively. Feng et al.~\cite{Feng:2013} evaluated the system on a dataset collected from 40 subjects. They have used three classifiers, DT, RF and BN. The best FAR was achieved by RF which was 8.93\% and the best FRR was achieved by BN which was 0.27\%. Clarke et al.~\cite{Clarke:2006} evaluated their system based on dataset of 30 samples collected from 30 subjects over three sessions. Their EER was 12.8\% on average which achieved by neural network classifiers. In contrast with all mentioned evaluation methods, Draffin et al.~\cite{Draffin:2014} evaluated their proposed system based on unconstrained dataset (i.e., they collected data overall application context without intervention or any supervision like what other studies did). The best result achieved by evaluating the system over input sessions of 15 keypresses with detection rate 67.7\%, where FAR was 14.0\% and FRR was 2.2\%, they built a discriminant algorithm based on neural networks.

\begin{table*}[h]
\centering
\scalebox{0.8}{
\begin{tabular} {l|p{1.8cm}| p{3.2cm} | p{3.8cm} | l | l} 
\toprule 
\textbf{Study} & \textbf{\# of Subjects}  & \textbf{Dataset} & \textbf{Features} & \textbf{Classification} & \textbf{Performance (\%)} \\ \hline
Li et al.~\cite{Li:2014}&22-76&MIT Reality&app name, Tel. number, cell, location, call (duration, time)&Neural Network& FRR:11.45, FAR:4.17\\
Kayacik et al.~\cite{kayacik:2014}&7,35,100&GCU,RiceLivelab, MIT Reality& + wifi, cpu load, light, noise ,magnetic field and rotation&PDF& DR: 53-99\\
Bassu et al.~\cite{bassu:2013}&NA&Private&app usage, time, location, HDI, bandwidth&Bayesian&NA\\
Gupta et al.~\cite{gupta:2012}&37-76&MIT Reality&GPS location, WIFI, bluetooth&Developed model& Precision:85, Recall:91\\
Shi et al.~\cite{Shi:2011}&50&-&SMS, Calls, Browser History, Location&-&-\\
\bottomrule
\end{tabular}}
\caption{Behavioral Profiling authentication methods. It contains the-state-of-the-art studies that conducted to develop touch-based active authentication systems. Number of subjects depicts how many participants contributed to the study. Dataset contains status of the collected data, private or public. Number of features that extracted from the collected data. Classifiers that used in the classification process. Finally, the best results that achieved by the developed method.}
\label{table:behavioral_profile}
\end{table*}
\subsection{Behavioral profiling}
Behavioral Profiling is the way in which the user interacts with the mobile sensors and services. Active authentication systems leverage these interactions to verify the user identity. It has been used for a long time to authenticate users based on their behavioral profile, where the literature of behavioral profiling was concentrated on network-based approaches such as user calling and service provider network to build a user profile~\cite{Hall:2005}. Also, some host-based approaches such as application usage and locations were used~\cite{Li:2011}. The intuition behind authentication system based on behavioral profiling is to build a profile of user activities over a period of time and compare that profile with the current user profile using some machine learning approaches.

Some recent active authentication techniques have been developed based on bahvioral profiling such as in~\cite{Li:2014, kayacik:2014, bassu:2013, gupta:2012}. Regarding these methods, we show what is the collected data and how they collect these data. Also, we discuss the feature extraction process in addition to the evaluation methods.

\subsubsection{Data collection}
One of the most common public dataset that have been used in the literature of behavioral profiling authentication is MIT Reality dataset~\cite{eagle:2006}. The MIT Reality dataset contains a rich amount of behavioral profiles sensors for 100 smartphone users from various departments of MIT. The data collected over the period of 9 months and contains sensor data such as call logs, bluetooth devices in proximity, cell tower IDs, application usage. Some proposed authentication system used MIT Reality dataset such as in~\cite{Li:2014, kayacik:2014, Shin:2012}. Other proposed systems have collected data by conducting a study. The evaluation results done by Li et al.~\cite{Li:2014} on the MIT Reality dataset achieved FRR of 11.45\% and FAR of 4.17\% overall the proposed framework. 

Kayacik et al.~\cite{kayacik:2014} proposed a data driven technique that compares the current user profile with the stored one. If the behaviour deviates sufficiently from the established norm, actions such as explicit authentication can be triggered. They evaluated the proposed system using three datasets, GCU, Rice Livelab and MIT Reality. GCU dataset consists of a collection from 7 staff and students of Glasgow Caledonian University. It was collected in 2013 from Android devices and contains sensor data from wifi networks, cell towers, application use, light and sound levels and device system stats. The duration of the data varies from 2 weeks to 14 weeks for different users. Rice Livelab~\cite{Shepard:2011} dataset was built over 35 users, all of them were students at Rice University or Houston Community College. The data was collected from iPhone 3GS devices between 2010 and 2011 and contains sensor data such as application use, wifi networks, cell tower IDs, GPS readings, battery usage and accelerometer output. The duration of the data varies from a few days to less than one year for different users.

Shi et al.~\cite{Shi:2011} developed a data collection application and posted it in Android marketplace. It has been downloaded by 276 users but only 50 users who kept it for a period of 12 days or more. They formulated their dataset based on the 50 users and they evaluated their proposed algorithm based on them. The dataset contains SMS, Phone calls, Browser history and Location. They used only two metrics to evaluate the proposed algorithm. First metric is the number of times the legitimate user used the device before a failed authentication. Second metric is the number of times the adversary used the device before detection.

Gupta et al.~\cite{gupta:2012} conducted several experiments using large-scale data collection tool~\cite{kiukkonen:2010}. They built a dataset that contains GPS location traces and regular scans of WiFi and Bluetooth radio environments of a large number of users. They developed contexts of interest (CoIs: a context that is significant to the user) identification algorithm. They evaluated their proposed prototype on the collected dataset which achieved Precision:85 and Recall:91.

Bassu et al.~\cite{bassu:2013} developed a new behavioral profiling authentication technique that combined four essential behavioral elements corresponding to what, where, when, and how. Apps usage constitutes the what, location and pace of movement defines where, clock time captures when, and gesture or input-output interactions captures how. They extracted set of features base on spatial and temporal analysis. Then they developed a classification model based on Bayesian classifier. Summary of the proposed behavioral profiling authentication methods is shown in Table~\ref{table:behavioral_profile}.

The main advantage of behavioral profiling biometric authentication is the capability of providing continuous and transparent authentication when users interact with their mobile devices, where all profile sensor data could be acquired continuously and without interrupting the user. However, a major weakness is the performance inconsistency when users interact with the mobile phones in an unusual way.

\begin{table*}[!h]
\centering
\scalebox{0.8}{
\begin{tabular} {l|p{1.65cm}| l | p{2cm} | p{2cm} | p{2.5cm}  } 
\toprule 
\textbf{Study} & \textbf{\# of Subjects}  & \textbf{Dataset(\# of sample)} & \textbf{Platform}  & \textbf{Sensors} & \textbf{Performance (\%)} \\ \hline
Neverova et al.~\cite{neverova:2016}&1500&Abacus(27.62 TB)&Android&Neural Network& -\\
Hoang et al.~\cite{hoang:2015}&38&Public~\cite{hoang:2013}&Android&sequence of prehensile movements& FAR:0, FRR$\approx$16.81\\
Juefei-Xu et al.~\cite{Juefei:2012}&36&166f(50.5m)&Android&Accelerometer and Gyroscope& VR: 99.4, FAR:0.1\\
Derawi et al.~\cite{Derawi:2010}&51&37&Android&Accelerometer (AK8976A)& EER:20\\
Mantyjarvi et al.~\cite{Mantyjarvi:2005}&36&-&108/20m&3-D accelerometer& EER:7\\
\bottomrule
\end{tabular}}
\caption{Gait based authentication methods. It contains the-state-of-the-art studies that conducted to develop gait-based active authentication systems. Number of subjects depicts how many participants contributed to the study. Dataset contains status of the collected data, private or public. The platform Operating System in where the study conducted. The sensors that have been used to collect the motion data. Finally, the best results that achieved by the developed method.}
\label{table:gait}
\end{table*}
\subsection{Gait recognition}
Gait based active authentication systems identify users based on the way in which the user walk. It is one of the rare biometric traits that can be used to recognize the people. With the diversity of built-in sensors in smartphones such as accelerometer and gyroscope made the development of authentication systems based on this trait feasible. 

Mantyjarvi et a.~\cite{Mantyjarvi:2005} proposed an implicit authentication biometric method based on gait dynamics. They collected three-dimensional movement data from 36 users via body worn 3-D accelerometer device, where the users walked about 20 meter in their normal, fast and slow walking speeds. The experimental results showed that the best EER was 7\% and achieved by means of a signal correlation method. Rather than using stand alone accelerometer devices, Derawi et al.~\cite{Derawi:2010} leverage the low-energy accelerometer sensor in the mobile device to collect accelerometer data from 51 subjects. They created a dataset of 37 meters walking distance with 40-50 samples per second for each of the three directions x, y and z. They evaluated their system based on the collected dataset and achieved EER 20\%. In addition to the accelerometer data, Juefei-Xu et al.~\cite{Juefei:2012} used gyroscope data to estimate the orientation of the phone in a user's pocket. They built a dataset with 36 subjects where they walked for 166 feet (50.5 meter). They extracted set of discriminative features based on Continuous Wavelet Transform (CWT)~\cite{lang:1998}. The best result achieved based on normal to normal pace which was 99.4\% for verification rate at 0.1\% for FAR.

Another novel idea has been proposed by Hoang et al.~\cite{hoang:2015} which verify the user via a stored key which is biometrically encrypted by gait templates collected from a mobile accelerometer. Also, they investigated the discriminability of sensor-based gait templates to construct an effective gait-based biometric crypto-system. They created a dataset from 38 participants using Google Nexus One device. They achieved zero FAR with approximately 16.18\% FRR.

In contrast with all previous studies, Neverova et al.~\cite{neverova:2016} created unsupervised and unconstrained dataset which was collected from approximately 1500 volunteers using LG Nexus 5 research phones as their primary devices on a daily basis manner. It is the largest study of its kind, where the motion data acquired from three sensors: accelerometer, gyroscope with 200 Hz sampling rate and magnetometer with 5 Hz sampling rate. They concentrated to authenticate users based on their natural kinematics, the motion patterns of human body. Their results demonstrated that human kinematics convey important information about user identity. Table~\ref{table:gait} summarizes the results for all proposed methods that built based on motion sensors.

Gait analysis relies on mobile inertial sensors such accelerometers and gyroscopes for authentication. These sensors are non-contact and non-obtrusive which could be used to design  authentication methods that resistant to spoofing attacks.

\begin{figure}
\centering
\includegraphics[scale=0.32]{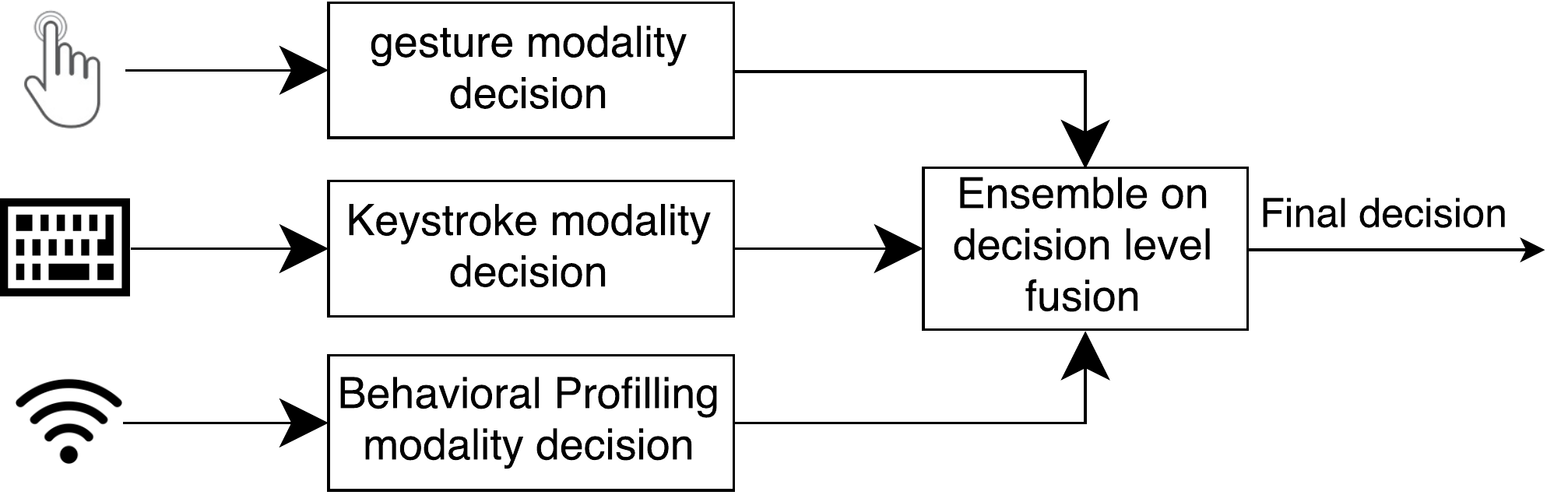}
\caption{Decision level fusion on multiple behavioral biometrics traits~(i.e., gesture, keystrokes and behavioral profiling)}
\label{fig:decision_level_multiple}
\end{figure}

\section{Fusing Different Biometric Traits}\label{fusion}
Fusing different behavioral biometric traits can improve the authentication accuracy and address some limitations and problems as in section~\ref{limitations}. There are different scenarios could be applied to perform the fusion which are as follows~\cite{Ross:2003, Ross:2004, Ross:2006, Faundez-Zanuy:2005, chen:2013}:
\begin{enumerate}
\item Sensor level fusion, which combine raw data that captured from different sensors for the same biometric traits.
\item Feature-level fusion, which combine different feature vectors that extracted from multiple biometric modalities in one new feature vector. 
\item Score level fusion, which apply the combination based on the matching score of each authentication modality. 
\item Decision level fusion, comprises decisions from multiple classifiers to make the final decision.
\end{enumerate}

Figure~\ref{fig:decision_level_multiple} illustrates the combination of multiple behavioral biometric traits based on decision level fusion where the local decisions~(i.e., calculated based on each biometric trait) are combined based on the majority voting method~\cite{Lam:1997}. The final decision~$D$ is predicted based on the \textit{plurality vote} of each local decision~$d_{ij}$ as follows~\cite{kuncheva2004combining}: 
\begin{equation}
D=\operatorname*{max}_j^C\sum_{i=1}^{L}d_{ij}
\end{equation}
where  \(j = \{1, \dots, C\} \) and $C$ represents the dimension of classes (i.e., in authentication case is 0 or 1 which means accept or reject), and \(i = \{1, \dots, L\} \) where $L$ represents the number of modalities' decisions (i.e., three modalities as shown in Figure~\ref{fig:decision_level_multiple}).

\section{Limitations and Challenges}\label{limitations}
Although there are several advantages associated with the active authentication systems, there are different limitations and problems are facing it which are as follows~\cite{Ross:2003, Ross:2004, Ross:2006, Khaleghi:2013, Serwadda:2016}:
\begin{itemize}
\item	\textbf{Noisy data}: the sensed data that recorded by the sensors devices that used in active authentication systems is always affected by some level of impreciseness in measurements.
\item	\textbf{Non-universality}: the active authentication system may not be able to collect meaningful data. In other words, the collected data might not reflect the correct user behavior.
\item	\textbf{Intra-class variations}: incorrect interaction with sensor, or the changing of the behavioral characteristics of the users at different time instances make variations.
\item	\textbf{Lack of uniqueness}: the interclass similarity between individuals will make some difficulties to differentiate between two users.
\item	\textbf{Vulnerabilities}: such as spoofing and robot attacks. for example Serwadda et al.~\cite{Serwadda:2016} developed two Lego-driven robotic attacks on touch-based authentication.
\end{itemize}

Although we could not cover all literature in active authentication but we could cover an important representative subset of the state-of-the-art methods. Based on these representative subset, there are some challenges and future trends that needs to be covered in the active authentication systems, which are as follows:
\begin{itemize}
\item	\textbf{Maximizing accuracy}: accuracy of active authentication system that built based on behavioral biometric traits is small. A better way to maximize the accuracy is needed.
\item	\textbf{Domain adaptation capability}: user characteristics change overtime. For instance, data collected in the enrollment phase may differ than those in the recognition phase. The active authentication system should use some domain adaptive method to handle this issue~\cite{Zhang:2015}.
\item	\textbf{Biometric feature extraction and selection}: feature extraction and selection are challenge process by nature. To extract and select an appropriate set of behavioral biometric features for active authentication system are more challenging. In depth analysis on the collected data is needed to select an appropriate set of behavioral biometric features.
\item	\textbf{Datasets}: one of weakest point in the majority of proposed active authentication systems is the lack of real-world datasets. Conducting a systematic user studies and experiments to get more users involved are very important factor to evaluate the active authentication system. The majority of active authentication system is evaluated on small datasets that contains hundreds of samples. Also, the availability of a public dataset for active authentication system research is needed~\cite{Crossler:2013}.
\item	\textbf{Usability}: the usability of active authentication systems is very important factor and needs to be handled~\cite{Adams:1999}.
\item	\textbf{Computation cost and energy consumption}: the capabilities of the smartphones are lower than the desktop systems. So the complexity cost should be considered in the design of active authentication systems on smartphones.
\end{itemize}

\section*{Acknowledgments}
We would like to thank our colleagues for their feedback on the earlier version of this Paper. The first author would like to thank Egyptian Mission sector for the doctoral scholarship.

\section*{References}
\bibliographystyle{elsarticle-num}
\bibliography{references}

\end{document}